\documentclass[journal=jacsat,manuscript=article]{achemso}

\usepackage{chemformula} 
\usepackage[T1]{fontenc} 

\usepackage{bm}
\usepackage{braket}
\renewcommand{\v}[1]{\bm{\mathrm{#1}}}

\usepackage{csquotes}
\MakeOuterQuote{"}

\author{E. I. Harris-Lee}
\affiliation[BigPharma]
{Max-Born-Institute for Non-Linear Optics and Short Pulse Spectroscopy, Max-Born Strasse 2A, 12489 Berlin, Germany}
\alsoaffiliation
{Max Planck Institute of Microstructure Physics, Weinberg 2, 06120 Halle, Germany}

\author{J.~K. Dewhurst}
\affiliation[BigPharma]
{Max Planck Institute of Microstructure Physics, Weinberg 2, 06120 Halle, Germany}

\author{W. Chen}
\affiliation[BigPharma]
{Max Planck Institute of Microstructure Physics, Weinberg 2, 06120 Halle, Germany}

\author{S. Hu}
\affiliation[BigPharma]
{Max-Born-Institute for Non-Linear Optics and Short Pulse Spectroscopy, Max-Born Strasse 2A, 12489 Berlin, Germany}

\author{Sam Shallcross}
\email{shallcross@mbi-berlin.de}
\affiliation[BigPharma]
{Max-Born-Institute for Non-Linear Optics and Short Pulse Spectroscopy, Max-Born Strasse 2A, 12489 Berlin, Germany}

\author{Sangeeta Sharma}
\affiliation[BigPharma]
{Max-Born-Institute for Non-Linear Optics and Short Pulse Spectroscopy, Max-Born Strasse 2A, 12489 Berlin, Germany}
\alsoaffiliation
{Institute for theoretical solid-state physics, Freie Universit\"at Berlin, Arnimallee 14, 14195 Berlin, Germany.}
\alsoaffiliation
{Halle-Berlin-Regensburg Cluster of Excellence CCE}

\title[An \textsf{achemso} demo]
{Ultrafast N\'eel vector switching}

\abbreviations{}

\begin{document}

\begin{tocentry}

\end{tocentry}

\begin{abstract}
We predict ultrafast switching in a chiral anti-ferromagnet that occurs at femtosecond times, nearly 5 orders of magnitude faster than the torque induced nanosecond switching previously observed. The physical mechanism, quite different from that which drives slow switching, involves the creation of massive effective magnetic fields by ultrafast spin current injection. Identifying these fields as key to femtosecond rotation, we establish simple practical rules for their maximisation with wide applicability to all magnetised materials. Employing state-of-the-art time-dependent density-functional theory and using the example of chiral magnet, Mn$_3$Sn, we induce ultrafast rotation enough to drive the switching of magnetic order between the six possible non-collinear ground states. We further demonstrate the possibility of undoing this switching by subsequent injection of oppositely polarized spin current. Our findings place chiral anti-ferromagnets as a materials platform for femtosecond N\'eel-vector switching, opening a route towards the manipulation of magnetic matter at ultrafast times.
\end{abstract}

\section{Introduction}

Control over magnetic order by ultrafast laser pulses represents a key route to utilizing condensed matter systems in information processing. Switching the spin moment orientation of ferromagnet, now established at sub-picosecond times \cite{igarashi_optically_2023,remy_accelerating_2023,harris-lee_spin_2024}, provides a basis for such light-wave manipulation. In the context of an ultrafast spintronics, however, anti-ferromagnets\cite{
baltz_antiferromagnetic_2018,
nemec_antiferromagnetic_2018} provide several advantages over ferromagnetic order. These include, the suppression of stray magnetic fields allowing for device component isolation, faster coherent switching, as well as the possibility of generation of pure spin currents (i.e. the transport of spin in the absence of charge flow), opening the way to a next generation energy efficient electronics.

Anti-ferromagnets have, furthermore, also been demonstrated to support repeated switching back and forth between distinct configurations, a crucial requirement for the efficient manipulation of magnetic order. Such materials thus present unique opportunities for the ultrafast control of magnetism by light \cite{thielemann-kuhn_ultrafast_2017,fabiani_supermagnonic_2021,
dannegger_ultrafast_2021,bossini_laser-driven_2019,nemec_antiferromagnetic_2018,
zhao_large_2021,kang_spin_2023,hamara_ultra-high_2024,zhang_spin_2024,Lee2025}.
However, while light induced current switching of ferromagnets has been demonstrated at sub-picosecond times \cite{igarashi_optically_2023,remy_accelerating_2023,harris-lee_spin_2024},  for \emph{chiral} anti-ferromagnets state-of-the-art switching times, of the order of nanoseconds, are nearly 5 orders of magnitude slower.

Here, using time-dependent density functional theory, we predict reversible switching between the degenerate configurations of a non-collinear magnetic order (Mn$_3$Sn) at ultrafast (100~fs) time scales. We show that a spin current, whose polarization is perpendicular to the in-plane compensated non-collinear magnetic order of Mn$_3$Sn, induces a continuous and ultrafast rotation of the chiral spin structure, allowing the system to be driven between the six distinct magnetic ground states that this material possesses.
Our work thus places chiral anti-ferromagnets as materials for which reversible switching of magnetic order can be achieved at femtosecond times, with concomitant rich possibilities for the manipulation of magnetic order at ultrafast times.
 
\section{Spin switching in Mn$_3$Sn}

In considering magnetic switching in a fully compensated (i.e. zero net moment) magnet it is important to address the challenge of probing, and therefore reading, a magnetic order parameter in the absence of global magnetisation.
Chiral anti-ferromagnets provide a solution: these have topologically non-trivial characteristics, leading to time-reveral-broken responses, such as the anomalous Hall effect and anomalous Nernst effect, which are readily measurable and can be used to determine the magnetic orientation.
Among the strongest of these effects is found in Mn$_3$Sn, which has an octupole order parameter $E_{1g} (T_x^\gamma)$ and thus time reversal symmetry breaking \cite{suzuki_cluster_2017}. It possesses an anomalous Hall effect both large and robust, even appearing in relatively disordered samples\cite{ikeda_anomalous_2018,tsai_electrical_2020}.
This is the material we will employ in the present study in a search for femtosecond scale switching of magnetic order.

\begin{figure}[t!]
\includegraphics[width=1\columnwidth, clip]{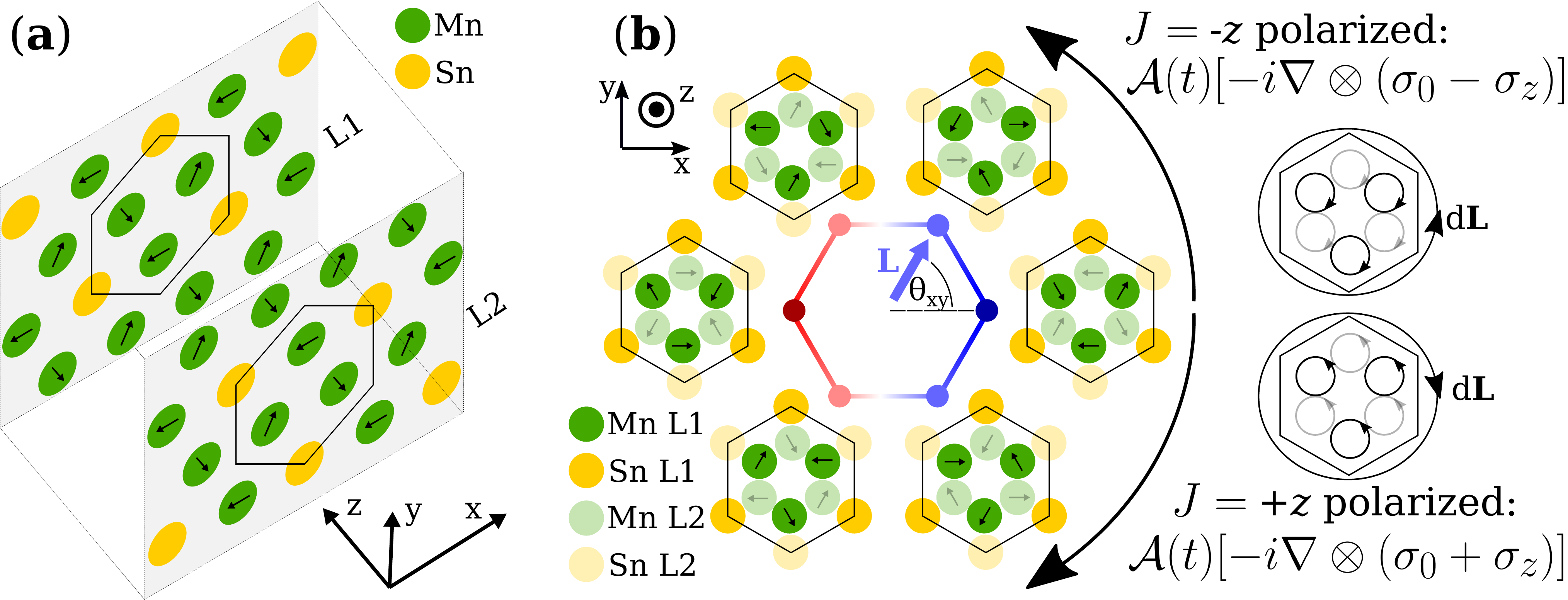}
\caption{
{\it Schematic illustration of Mn$_3$Sn lattice and magnetic moments.}
(a) Primitive unit cell of Mn$_3$Sn, with boundaries indicated by solid black lines. The Mn sites form a kagome pattern with two layers, denoted L1 and L2.
(b) The six ground state magnetic configurations, each separated by $\pi/3$ rotation of the local moments, with the magnetic ordering vector, $\mathbf{L}$.
}\label{fig1}
\end{figure}

In Fig.~\ref{fig1}(a) is shown schematically the crystal structure and magnetic ground state of Mn$_3$Sn along with, panel (b), the six distinct non-collinear ground states with the associated N\'eel vector (defined here not by a 2 sublattice difference).
Switching between these non-collinear ground states is facilitated by small in-plane magnetic anisotropy, and indeed recent work has shown that Mn$_3$Sn can be switched via spin-orbit torque \cite{tsai_electrical_2020,takeuchi_chiral-spin_2021,pal_setting_2022}. In all these investigations, however, torque induced nanosecond timescale switching is achieved, which is four orders of magnitude slower than the femtosecond times achieved for ferromagnets \cite{igarashi_optically_2023}.
    
\section{Method}

To investigate numerically the possibility of ultrafast switching in this material we employ time-dependent density functional theory (TD-DFT) \cite{RG1984,my-book,sharma2011,sharma_computational_2022}, a first principles framework that has proven to be especially useful for predicting the dynamics of spins in the femtosecond regime \cite{willems_optical_2020,dewhurst2020,hofherr2020}.
Within this framework the equation of motion is the time-dependent Schr\"odinger equation, 
\begin{eqnarray}
\label{eq:fullksham}
i\frac{\partial \phi_{n\v k}(\v r,t)}{\partial t}=\left[
\frac{1}{2}\left(-i{\boldsymbol \nabla} -\frac{1}{c}{\bf A}(t)\right)^2 \right. \nonumber\\
+v_{\rm s}(\v r,t) + \frac{1}{2c} {\boldsymbol \sigma}\cdot{\bf B}_{\rm xc}(\v r,t) \\\nonumber
+\left.
\frac{1}{4c^2} {\boldsymbol \sigma}\cdot ({\boldsymbol \nabla}v_{\rm s}(\v r,t) \times -i{\boldsymbol \nabla})\right]\phi_{n\v k}(\v r,t)
\end{eqnarray}
where $\phi_{n\v k}(\v r,t)$ are time-dependent Kohn-Sham spinor orbitals with quasi-momentum $\v k$ and state index $n$, and the usual TD-DFT Hamiltonian is contained in square brackets.
In this Hamiltonian, ${\bf A}(t)$ represents the time-dependent vector potential, $\boldsymbol \sigma$ is the Pauli matrix vector, and $v_{\rm s}(\v r,t)=v_{\rm ext}(\v r)+v_{\rm H}(\v r,t)+v_{\rm xc}(\v r,t)$ is the Kohn-Sham effective scalar potential. The electron-nuclei interaction, which completely characterises a material, is included through the external potential, $v_{\rm ext}(\v r)$, while
the electron-electron interaction is split into the Hartree potential, $v_{\rm H}(\v r,t)$, which is the exact classical part, and the exchange-correlation (xc) terms: the scalar potential $v_{\rm xc}(\v r,t)$ and the xc magnetic field, ${\bf B}_{\rm xc}(\v r,t)$.
The time dependent magnetization density, $\mathbf{m}(\mathbf{r},t)$, the key quantity for investigating early time laser induced changes in magnetic order, may be obtained directly from the expectation of the spin operator, $\boldsymbol{\sigma}$,
\begin{equation}
\mathbf{m}(\mathbf{r},t) = \sum_{n \v k}^{\rm occ.} \phi_{n \v k}^*(\mathbf{r},t) \ \boldsymbol{\sigma} \ \phi_{n \v k}(\mathbf{r},t),
\label{eqn:expectation}
\end{equation}
where $\psi_n$ is the spinor with index $n$ obtained from solution of the TD-DFT Hamiltonian, Eq.~\ref{eq:fullksham}.

The generation of laser induced ultrafast spin currents in experiment typically involves light generating a spin current in one part of a device, that then is injected into the magnetic material to be studied. Such complex heterostructures, with present day computer power, cannot be treated within TD-DFT. Recently, it was shown how TD-DFT can be extended to include spin-dependent vector potentials in the system Hamiltonian\cite{harris-lee_spin_2024}, such that the physics of magnetic changes driven by spin current can be simulated without the need for calculating the full device structure. This proceeds by the addition to the TD-DFT Hamiltonian of the term
\begin{equation}
     -\frac{i}{c} \sum_{ j \mu} {\mathcal{A}}_{ j \mu}(t) \nabla_{ j} \sigma_{ \mu},
    \label{sp}
\end{equation}
where $\frac{1}{c}$ is the coupling constant, $j$ are Cartesian indices, $\mu$ are Cartesian indices plus a $0$th index to include the standard $U(1)$ vector potential, $A$, in this more general expression, and these indices specify components of the linear momentum $-i{\nabla}$ and spin vector operator $\bm{\sigma}$ respectively.
These are needed to form the conjugate operator to the conserved spin current density, characterised by the tensor product $-i{\nabla} \otimes \bm{\sigma}$.
More detail, including a general form given to the matrix functions $\mathcal{A}_{\rm j \mu}(t)$, is given in Ref.~\cite{harris-lee_spin_2024}.
The form allows for arbitrary matrix functions $\mathcal{A}_{ j \mu}(t)$ and thus the simulation of any spin and charge current, including both the limits of pure charge current, and pure spin current.
Since we are concerned with a sub-system of a wider device, we do not add terms to make the Hamiltonian covariant, and we focus on paramagnetic current densities, for more details see the extended Method SI~1 in the Supplemental document. 

Our method is implemented in the state-of-the-art version of the open source Elk code \cite{elk,dewhurst2016}. We use the fully non-collinear spin-dependent version \cite{krieger2015,dewhurst2016} with the adiabatic local spin density approximation for the exchange-correlation potential. Full numerical details may be found in the Supplemental document SI~0.

\section{Results}
 
\begin{figure*}[t!]
\includegraphics[width=1\columnwidth, clip]{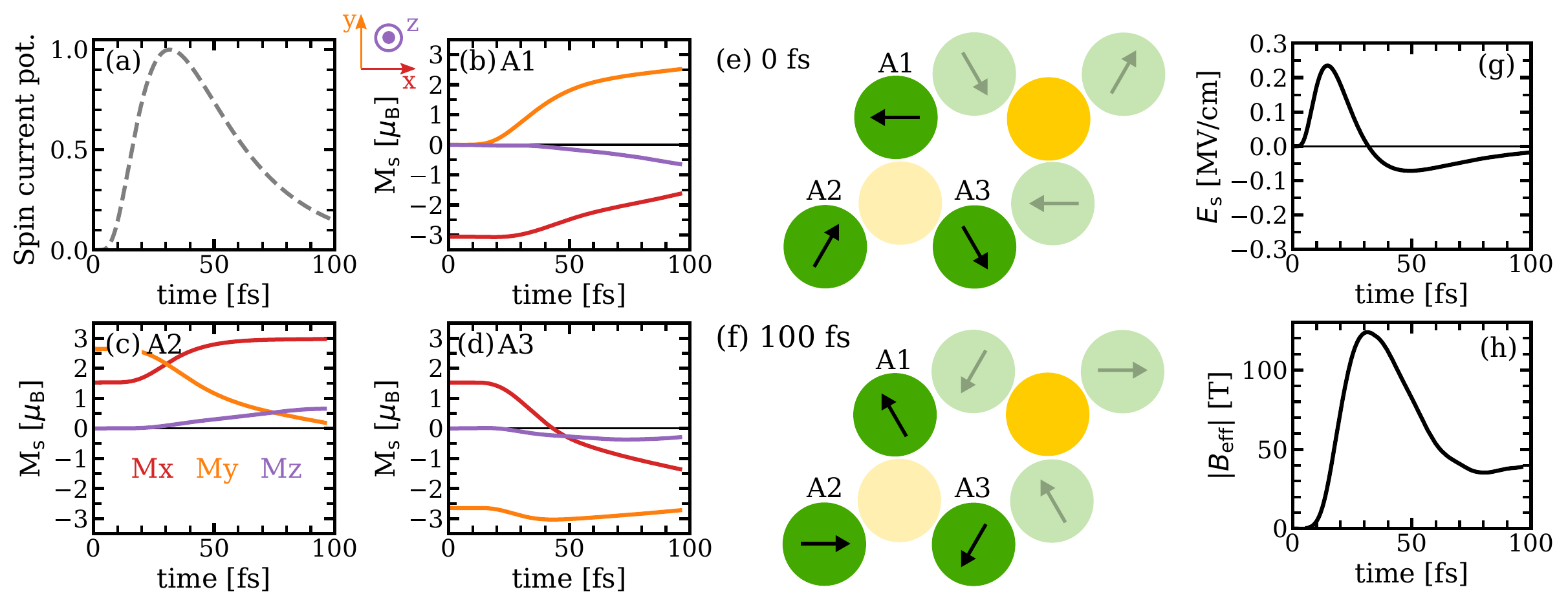}
\caption{
{\it Ultrafast rotation of local moments by spin current.} 
(a) Time-dependent amplitude of spin current potential, with polarisation vector out of plane.
(b,c,d) The calculated time evolution of the magnetisation vector, shown for each of the three distinct sites.
Note that a small out-of-plane ($m_{\rm z}$) magnetization develops.
Initial, (e), and final, (f), local moment vectors; in each case the second inequivalent layers is shown as and the faded sites. After the spin current the local moments are rotated approximately 60 degrees.
(g) Effective spin-dependent electric field for spin current potential shown in (a).
(h) Effective magnetic field, calculated from in-kagome-plane magnetisation dynamics.
}\label{fig2}
\end{figure*}

We apply a spin polarised current potential with time-dependence shown in Fig.~\ref{fig2}(a), where the amplitude and width of the potential are chosen to fit those for spin currents which can be stimulated using ultrafast laser pumps \cite{bierhance_spin-voltage-driven_2022,rouzegar_broadband_2023}.
This spin current is polarised perpendicular ($z$) to the Kagome plane ($xy$) of the local moments. Evidently, this spin current generates a pronounced change in all spin moments, Fig.~\ref{fig2}~(b-d), with the primary change a rotation around the axis of the spin current polarisation, but with also a slight increase in magnetisation in the $z$ direction.

The direction of each moment at the start and end of the calculation are shown explicitly in Fig.~\ref{fig2}(b-d), from which it can be seen that the overall effect of the spin current is to generate a 60 degree rotation of each of the local moments.
Crucially, as magneto-crystalline anisotropy dictates  an energy minima every 60 degrees, the overall result is a rotation of the N\'eel vector between the inequivalent non-collinear ground states, i.e. order parameter has been successfully switched.

The time scale of this switching is significantly shorter than the limit for precessional mechanisms such as spin transfer torque or spin orbit (SO) torque, and thus the question now arises as to the underlying mechanism allowing this ultrafast N\'eel vector control.
We first consider the role of spin-orbit coupling:
this represents one of the dominant interactions in femtoscale spin dynamics responsible, for example, for ultrafast light induced demagnetization. Switching off the spin-orbit term in the TD-DFT Hamiltonian, however, results in almost no change in the spin dynamics, ruling out SO torques as playing the decisive role.
Nor are the spin dynamics inertia-driven \cite{kimel_inertia-driven_2009}, given that the most significant changes occur during the driving field;
The spin dynamics must, therefore, be directly driven by the spin current itself.

It is thus natural to consider that the spin current is generating an effective transient magnetic field, and to probe this we attempt to fit the dynamics to those that would be induced in such a case. To facilitate this we considered until this point the dynamics of the local moments $\v m(t)$ obtained by integrating the TD-DFT spin density in muffin-tin spheres that circumscribe the Mn sites. In fact, the dynamics of the full vector field $\v m(\v r,t)$ can be revealed, see Supplemental~2. We thus attempt to fit the local dynamics to those governed by an effective magnetic field:
\begin{equation}
\frac{d\v m(t)}{dt} =  \alpha  \v B(t) \times \v m(t)
\label{eqn:LL}
\end{equation}
i.e. the form of the magnetic-field driven part of the spin dynamics derived through Ehrenfest's theorem \cite{dewhurst2018} (or of the low damping Landau-Lifshitz dynamics). We find that the TD-DFT spin dynamics can be almost perfectly reproduced by Eq.~\ref{eqn:LL}, and in this way calculate the effective magnetic field which is required to drive the spin rotation seen in TD-DFT, Fig.~\ref{fig2}(h). The spin current thus corresponds to the effect of  magnetic field, of the order of 100~T which is transient and only lasts a short (femtosecond) time, however, is enough to rotate the spins.

At this point we emphasise that this massive effective magnetic field arises from a magnitude of ultrafast current density which is chosen in line with expected currents via ultrafast laser pulses, $3.3\times10^{15}$~Am$^{-2}$ (1.1~Nm$^{-1}$). While this is certainly high compared to typical spintronic switching values, it is only necessary for 50 femtoseconds or less. We also calculate the spin dependent electric field from the spin current potential, $ E_{j \mu}(t)=-\frac{1}{c}\frac{d \mathcal{A}_{j \mu}(t)}{dt}$. This is shown in Fig.~\ref{fig2}(g), and we find it comparable to electric fields known to be created by spin currents in experiment, see Ref.~\cite{rouzegar_broadband_2023} for example.

The small increase in out-of-plane polarization, 0.5 $\mu_{\rm B}$ per atom, induced during this ultrafast rotation of the N\'eel vector representing non-collinear order suggests that application of an oppositely polarized spin current will undo the rotation induced by the first current pulse. This indeed is the case and, as shown in the Supplemental SI~3, multiple back and forth switching events can be supported in the early time regime. We thus conclude that transient spin currents represent a route to full control over the the N\'eel vector of Mn$_3$Sn.

We now consider in detail both the dependence of the rotation rate on the amplitude of the spin current, and, moreover, the type of spin current and the robustness to current polarization, an important question for experimental realization of this effect.
It is important to note that the following rules apply to any magnetized material.
From this point any reference to `spin current' is meant in the literal sense, not a contraction of `spin polarised charge current' which is commonly referred to simply as `spin current' in literature.

\begin{figure}[t!]
\includegraphics[width=0.98\columnwidth, clip]{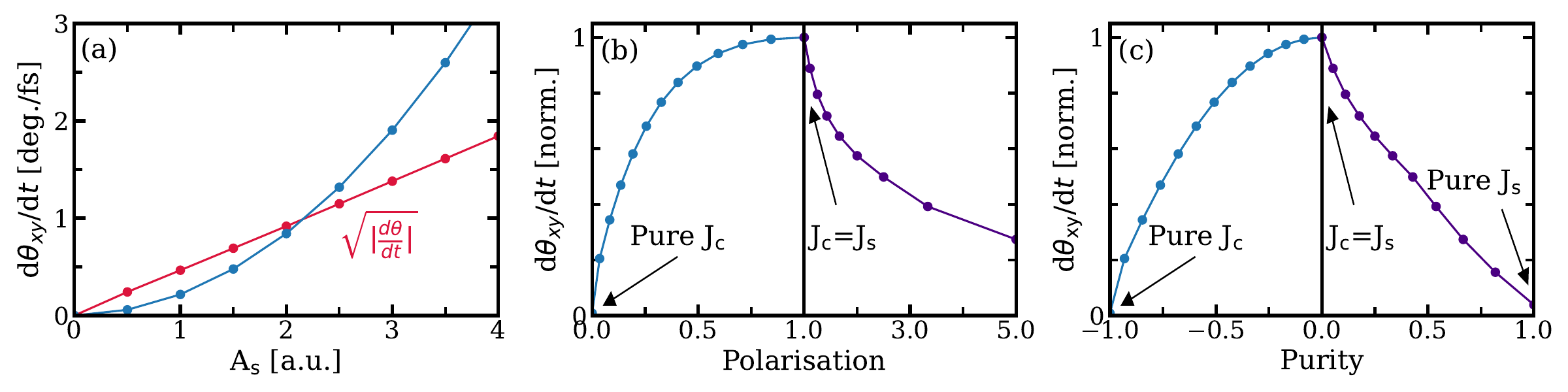}
\caption{
{\it Dependence of rotation on current amplitude and polarization.} 
(a) Rate of rotation versus spin current potential  (over fixed duration). The relationship is quadratic, as made clear by the plot of the square root of the rotation rate which is visibly linear.
(b) Rate of rotation versus polarisation ($P$) of the spin current potential. We define polarization as $P=\frac{\mathcal{A}^{\rm +z}-\mathcal{A}^{\rm -z}}{\mathcal{A}^{\rm +z}+\mathcal{A}^{\rm -z}}\approx \frac{J^{\rm +z}-J^{\rm -z}}{J^{\rm +z}+J^{\rm -z}}$, where ${\pm z}$ indicates spin polarization perpendicular to the kagome plane such that $J^{\rm +z}$ is the current density of ${\rm +z}$ spin electrons and $\mathcal{A}$ is a current potential.
For polarisation up to 1 $\mathcal{J}^{-z}$ is varied for fixed $\mathcal{J}^{+z}$ while for polarisation above 1 the total spin current is fixed while the charge current is varied.
The rate of rotation is highest for equal charge and spin current ($P=1.0$), and it goes to zero for pure charge currents ($P=0.0$) and for pure spin currents ($P=\infty$, not shown).
(c) Rate of rotation versus `spin purity' $=(\mathcal{A}-A)/(\mathcal{A}+A)$. 
}\label{fig3}
\end{figure}

For the aformentioned spin polarised charge current, the instantaneous rate of rotation depends, as may be seen in Fig.~\ref{fig3}(a), quadratically on the amplitude. This in turn implies that the effective magnetic field depends quadratically on the amplitude. Evidently, this dramatically reduces the amplitude of the spin current required to generate femtosecond scale rotation. 
This quadratic behaviour can be broken down further, since it can be shown that the rotational magnetisation dynamics follow
\begin{equation}
\frac{d\mathbf{m}({\v r},t)}{dt} \propto \sum_j A_j(t) \mathcal{A}_{ j}(t) \times \mathbf{m}(\mathbf{r},t),
\label{eqn:LL2}
\end{equation}
where this has the form of magnetic field driven dynamics, Eq.~\ref{eqn:LL}, but replacing magnetic field by charge and spin current potentials.
Thus to a good approximation
\begin{equation}
\frac{d\mathbf{m}({\v r},t)}{dt} \propto (\mathbf{j}(t) \cdot \mathcal{J}(t)) \times \mathbf{m}(\mathbf{r},t),
\label{LL2}
\end{equation}
As a higher spin polarized current requires both higher charge current $\mathbf{j}$ \emph{and} spin current $\mathcal{J}$, the origin of the quadratic behaviour is revealed.
We emphasise however that the rotation rate can also be increased by raising each of these currents individually: higher spin current leads to a greater rate of spin rotation, but so does higher charge current (as long as spin current is non-zero).
The counterintuitive, yet crucial, corollary to this point is that non-zero charge current is required to achieve ultrafast rotation: a perfectly pure spin current is ineffective.

This quadratic relationship is also reflected in the polarisation dependence of the rotation rate, shown in Fig.~\ref{fig3}(b).
Clearly, no rotation is stimulated by an spin unpolarised charge current. Increasing the spin current then results in a rapid increase in the rate of rotation, reaching 80\% of the maximum, which occurs for 100\% spin polarization, at only 35\% polarisation. The rotation rate is thus highly robust to degradation of the spin current from full 100\% polarization, a key point for experimental implementation as a perfect spin polarization source is unlikely, with further degradation over time due to spin-orbit coupling.
Moreover, a current density of only 1\% polarisation ($3.3\times10^{16}$~Am$^{-2}$ and 0.1~Nm$^{-1}$) can still correspond to a rotation rate in excess of 1$^{o}$/fs.

However, as the polarisation exceeds unity, representing spin transport in excess of the total charge transport the rate of rotation again begins to drop. This can be seen in Fig.~\ref{fig3}(c) in which we plot the spin purity, interpolating between the limits of a pure charge current (spin purity -1) and a pure spin current  (spin purity unity i.e. the flow of spin without charge). In both these limits the rotation rate falls to zero.
(Note that the discontinuity at polarization $P=1.0$ and purity $0.0$ arises due to the fact that for $P<1.0$ the $-z$ spin current is decreased for fixed $+z$ spin current, until 100\% polarized current, and thereafter the spin current is held fixed and the charge current decreased.)

\section{Discussion}

Reversible switching of the chiral order of Mn$_3$Sn can be driven by successive pulses of transient spin current, with considerable robustness to degradation from the 100\% spin polarized current for which the effect is most efficient. Switching on 100~fs times is possible even with 1\% polarization, if the charge current is high enough. The rotation of non-collinear structure is generated by a massive transient effective magnetic field created by transient currents. In contrast to near equilibrium manipulation, the transient nature of the effective magnetic field allows a massive short field burst able to drive an ultrafast rotation.

We emphasise that this mechanism for switching is not unique to Mn$_3$Sn or non-colinear magnets, see SI~5, though Mn$_3$Sn does have several advantageous properties:
(i) switching of the order parameter requires only a rotation of $\pi/3$ between the 6 inequivalent ground states, not the much larger angle $\pi$ required in collinear systems;
(ii) relatively long spin coherence length \cite{Lee2025} and low out-of-plane spin accumulation;
(iii) large and robust anomalous Hall effect and chiral magnetic order which offer routes to read out the magnetic structure.
Perhaps most importantly, Mn$_3$Sn, with its low stray field, can be placed particularly near to a spin current source. 

In summary, we find a preponderance of evidence that a substantial rotation -- enough to switch the magnetic order parameter -- ought to be achievable in the femtosecond limit. This is made possible by the rapid quadratic rise of the magnetic-field-like strength of intense ultrafast spin polarised currents. On the other hand, pure types of spin current are ineffective. Further investigation and direct experimental confirmation of this effect could set the stage for even greater device functionality in the femtosecond time domain.

\begin{acknowledgement}

E. I. Harris-Lee would like to thank the DFG for funding through project SH498-8/1. We thank the DFG for funding through Project 328545488 TRR227 (project A04). S. Shallcross and S. Sharma would like to thank Leibniz Professorinnen Program (SAW P118/2021) and German Excellence Strategy – EXC3112/1 –533767171 (Center for Chiral Electronics) for funding. Computations were performed on the HPC Systems at the Max Planck Computing and Data Facility. The authors acknowledge the North-German Supercomputing Alliance (HLRN) for providing HPC resources that have contributed to the research results reported in this paper.

\end{acknowledgement}

\begin{suppinfo}

Supporting information is available:
\begin{itemize}
  \item Filename: SI.pdf -- this document contains supplemental calculations and associated discussion referred to in the text but not elaborated upon.
\end{itemize}

\end{suppinfo}

\clearpage

\bibliography{Ledge}

\end{document}